\documentclass{elsart}
\usepackage{graphics}
\usepackage{amssymb}
\usepackage{epsfig}
\begin{document}

\begin{frontmatter}

\title{Decay of Affleck-Dine Condensates with Application to Q-balls} 
\author[andy]{Andrew Pawl  }

\address[andy]{Michigan Center for Theoretical Physics, Randall Laboratory,
University of Michigan, 500 E. University Ave., Ann Arbor, MI 48109, USA}

\begin{abstract}
	Analytical and numerical estimates show that a charged Affleck-Dine
condensate will fracture into Q-balls only when the Hubble time is significantly
larger than the inverse soft-breaking mass of the field in question.  This would
generally imply that the decay of the field into light fermions will compete
with Q-ball formation.  We will show that for typical flat directions
the large field value will significantly suppress decays of the condensate
to fermions even if no baryon charge asymmetry exists.
We will consider the details of the
decay process for a condensate that does carry charge, and show that
it is qualitatively different from that of an uncharged condensate. 
Finally,
we will consider the possibility of resonant production of heavy bosons.
We will show that this can have a strong effect on the condensate.  
Contrary to intuition, however, our results indicate that boson production
would actually assist Q-ball formation in condensates with significant charge.

\end{abstract}

\end{frontmatter}

\section{Introduction}

A Q-ball is a non-perturbative solution of the equation of motion
for a scalar field which is charged under a continuous U(1) symmetry
\cite{coleman1}.  The MSSM requires several such scalars, and so
it is expected that Q-balls could be formed in a supersymmetric
universe \cite{kusenko1}.  In fact, Q-balls seem inevitable in
the context of Affleck-Dine \cite{AD} (AD) baryogenesis.  Here, a flat direction
composed of several squark fields gains a large expectation value
and is set into coherent rotation by the action of a phase-dependent
term in the potential.  Such a charged scalar condensate has been shown
to fracture into Q-balls in analytical
treatments \cite{shaposhnikov,EM,fujii} and in numerical simulations
\cite{KK}.  

Both numerical and analytical estimates agree that in gravity-mediated
SUSY breaking models, Q-balls will only form when the age of the universe
has reached a value of order $\sim 10^{3} m_{\phi}^{-1}$ where $m_{\phi}$
is the soft-breaking mass of the AD condensate field (assumed to be of
order 1 TeV).  Naively speaking, however, we would estimate the decay
width of the scalar field into light fermions to be of order:
\begin{equation}
	\Gamma(\phi \rightarrow \psi \psi) \approx \frac{g^{2}m_{\phi}}
{8\pi}
\end{equation}
where we have substituted the gauge coupling $g$ for the usual Yukawa
coupling because the squark fields making up $\phi$ are coupled to
gluino/chargino plus quark through gauge interactions.  

If we do not invent a suppression for
this decay by making the fermions heavy then we see 
that the time for our condensate to decay can be as short as order 
$100 m_{\phi}^{-1}$.  Thus, we expect a competition between decay of
the condensate into light fermions and fracturing of the condensate
into Q-balls.  
As we will explore in this paper, however, there is not only
a suppression
to the decays, but there is also an important difference between
the decay of a partially charge-asymmetric condensate and a
neutral one.  Taken together, these details
will ensure that Q-ball formation is uninterrupted (and perhaps
even aided) by decay 
of the condensate.

\section{Decay into Fermions}
\label{sec:imtime}

\subsection{Necessity of Non-perturbative Approach}

The physics leading to the suppression of decay into 
fermions is familiar.  Any fermions coupling directly to the $\phi$
field will gain a mass of order $g|\phi|$ where $g$ is the relevant
coupling and $|\phi|$ is the magnitude of the complex scalar field.
In modern formulations of the AD scenario \cite{randall}, when the
Hubble constant $H$ reaches $m_{\phi}$ the 
AD field will have a magnitude of order:
\begin{equation}
\label{eq:phivev}
        |\phi| \sim \left(
        m_{\phi} M^{n-3}\right)^{1/(n-2)}
\end{equation}
where $M$ is a large mass scale (order $M_{GUT}$ or $M_{pl}$) and
$n$ describes the flatness of the flat direction (two
standard choices are $n=4$ or $n=6$).  We will
take $n=6$ as our canonical value because it minimizes thermal
concerns \cite{campbell}.  For typical numbers, then, we 
expect $|\phi| > 10^{12} m_{\phi}$ (or $10^{9} m_{\phi}$
for $n=4$) so that all fermions
have effective masses much larger than the mass of the $\phi$
field.  

At first glance, this solves our problem completely since
the $\phi$ field is stable.  However, it is important that
when $H \leq m_{\phi}$ the $\phi$ field will begin to execute 
harmonic oscillations about $\phi=0$ \cite{randall}.  
This will result in a sinusoidally varying mass for the fields
coupled to the $\phi$ field.  Such a situation has already
been studied for real (non-complex) scalar condensates in the context of
post-inflation reheating, and has been shown to lead to
decay of the condensate \cite{dolgov1,traschen}.
We wish to expand this analysis
to the case of an oscillating complex $\phi$ field.  

Explicitly, we anticipate that after a few oscillations 
the $\phi$ field will see the effective potential:
\begin{equation}
	U(|\phi|) \approx m_{\phi}^{2} |\phi|^{2}
\end{equation}
where higher order terms can be neglected due to the
small size of $|\phi|$ \cite{randall}.  We can now make
the analogy between $|\phi|$ and the radial position $r$
of a particle in an $r^{2}$ potential.  We know from
basic classical mechanics that the angular momentum in
such a system will be conserved (in this case, ``angular
momentum'' is equivalent to baryon number), and further that the
particle will follow closed orbits \cite{gold}.  Numerical
integration shows that this approximation is very nearly
exact even in the presence of small corrections
due to non-renormalizable
terms and log running of the mass parameter.  

All of this amounts to the fact that we lose no generality in
parameterizing the final solution for the $\phi$ field in
the form of an ellipse centered on the origin:
\begin{equation}
\label{eq:ellipparam}
	\phi = a \sin(m_{\phi}t) - ib \cos(m_{\phi}t)
\end{equation}
where we have assumed that $a \geq b$.

Using the Noether current expression for the global U(1)
yields the expression for net baryon number density
of the condensate:
\begin{equation}
    n_{B} =  i \beta \left(\dot{\phi}^{*}\phi-\dot{\phi}\phi^{*}\right)
        = 2 \beta m_{\phi} ab
\end{equation}
where $\beta$ is the baryon charge per $\phi$ particle 
(usually 1/3).
Then, by assuming that each scalar particle associated with
the $\phi$ field has
an energy of approximately $m_{\phi}$, it is
simple to show that the ratio of net baryons to total scalars is:
\begin{equation}
\label{eq:chgfrc}
        \frac{ n_{B}}{n_{\phi}} = \frac{2\beta ba}{a^{2}+b^{2}}.
\end{equation}

From this expression, we can see that the limit $b =0$
corresponds to a completely uncharged scalar field.  The
limit $b = a$ indicates a total charge asymmetry (the
condensate is made up entirely of baryons or entirely
of antibaryons).  Intermediate values of $b$ indicate a
partially charge-asymmetric condensate.

Using the parameterization (\ref{eq:ellipparam}), we also find that:
\begin{equation}
\label{eq:param}
	|\phi| = \sqrt{b^{2} + (a^{2}-b^{2}) \sin^{2}(m_{\phi}t)}.
\end{equation}
Thus, we do anticipate an oscillating mass for any fields coupled
to the AD scalar.

\subsection{Method of Imaginary Time}

The method of imaginary time is convenient for the calculation of
non-perturbative production of fermions.  We will use the
results of \cite{dolgov1} essentially verbatim.
For an introduction
to the imaginary time formalism, see the review \cite{dolgov2}
and the references therein.

The method relies on finding the branch points
in the complex-time plane of the fermion Hamiltonian.  For
our fermions, we expect:
\begin{equation}
 	\mathcal{H} = \left(p^{2}+g^{2}|\phi|^{2}\right)^{1/2} = 
	\left(p^{2}+g^{2}b^{2}+g^{2}(a^{2}-b^{2})
	\sin^{2}(m_{\phi}t)\right)^{1/2}.
\end{equation}
Fortunately, exactly this form was treated in \cite{dolgov1}.

Two limits are analytically approachable using their results.  
First, note that by Equation (\ref{eq:phivev}) we always expect
$m_{\phi} \ll a$.  
Now, if we also
assume $b \ll a$ (equivalent to  $m_{0} \ll m_{1}$ in the conventions of
\cite{dolgov1}) we can take an analytical
limit: 
\begin{equation}
	\Gamma_{\phi} \approx \frac{e^{-\pi/2}m_{\phi}^{3/2}}{8\pi^{2}(ga)^{1/2}}
\end{equation}
(note that $e$ here is the base of the natural log).  In this
approximation we have basically followed \cite{dolgov2} exactly,
except that we have added the assumption that $\log(4 a/b)$
is of the order $\pi$.  Note that here we have assumed $b$ is
nonzero, and larger than the momentum $p$.  If $b$ were to
approach zero, the results of \cite{dolgov1,dolgov2} apply
exactly (there is no divergence).
Putting in typical numbers will tell us that
$\Gamma$ is of order $10^{-6} m_{\phi}$ at the largest ($10^{-5} m_{\phi}$
for $n=4$).  This
suppression is enough to keep our condensate intact until
Q-balls can form, even for the less-favorable $n=4$ case.

 Next let us take the limit $b \approx a$.  Here again we
can analytically approximate the
decay width (this is equivalent to $m_{0} \gg m_{1}$ in 
\cite{dolgov1}):
\begin{equation}
\label{eq:largeb}
        \Gamma_{\phi} \approx \frac{m_{\phi}^{3/2}}{16 \pi^{2} (ga)^{1/2}}
        \exp\left(-2 \frac{ga}{m_{\phi}}\ln\left[\frac{16 a^{2}}{a^{2}-b^{2}}
	\right]\right)
\end{equation}
which, for our typical numbers yields a decay width that has been
exponentially suppressed to the extent that it is effectively zero.

\begin{figure}
\begin{center}
\psfig{file=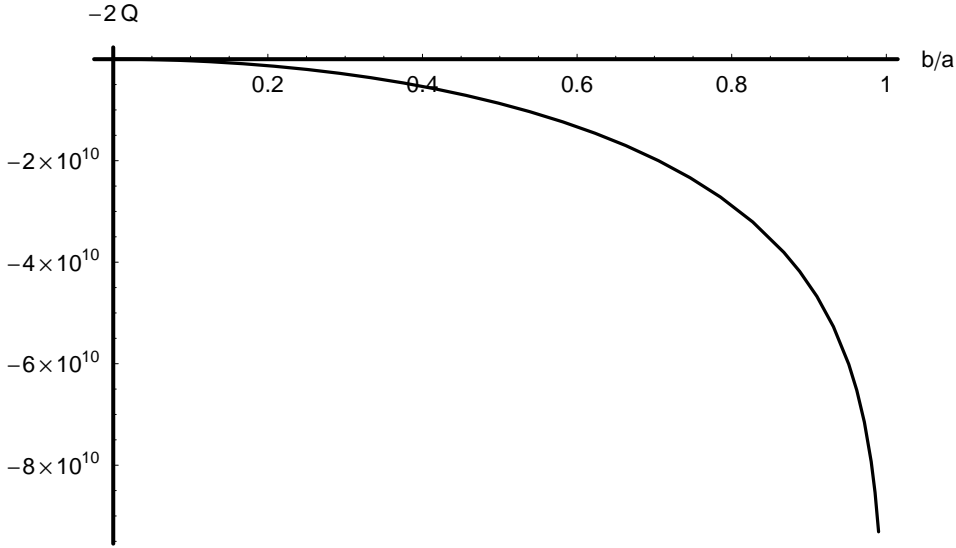,width=5.0in}
\caption{Exponential damping constant in the $\phi \rightarrow \psi\psi$
decay width for various values of $b/a$.  $\Gamma_{\phi} \sim e^{-2Q}$.
We have assumed $ga = 10^{10} m_{\phi}$.}
\label{fig:interp}
\end{center}
\end{figure}

In each of these limits for the ratio of $b$ to $a$, 
order of magnitude estimates show that
the decay rate is highly suppressed.  It is logical that the rate
does not peak between, but rather goes smoothly from one to the other.
To be safe, however, we can numerically evaluate the complete elliptical
functions outside the range of validity of our approximations to
estimate the exponential part of the suppression for arbitrary values
of $b/a$.  The
curve shown in Figure \ref{fig:interp} computes only the exponential
suppression of the decay rate, but it is enough to show
that this assumption
is correct.  
Thus, the large field values suppress the decay into fermions sufficiently
to allow Q-ball formation to proceed.

It is important to notice in Equation (\ref{eq:largeb}) that at $b=a$
the decay rate goes to zero.  This is no accident.  A completely charge
asymmetric scalar condensate is stable against decay into fermions --
even non-perturbative decay \cite{coleman2}.  This is an important
component of
Q-ball stability.

\section{Parametric Resonance}

Finally, we must consider the decay of our field into bosons.  
This decay will be suppressed by the same physics that 
affected the decay into fermions.
It is well known, however, that in the case of decay to bosons enhancement
from particle statistics can offset this suppression.  This phenomenon
was mentioned in \cite{dolgov1,traschen}.  The first modern treatment,
however, was \cite{linde1}.  We will follow the expanded version 
presented in \cite{linde2}.

We expect the $\phi$ field to have several couplings of the form
$g^{2}\phi|^{2}|\chi|^{2}$ where 
$g$ is the strong gauge coupling and $\chi$ is a scalar (a linear
combination of squarks that is orthogonal to the combination that
makes up the 
$\phi$ direction).  

This contribution to the Lagrangian will result in an oscillating
mass term for the $\chi$ field when the  
$\phi$ field is undergoing elliptical oscillation.  The
scalar $\chi$ will have the following equation of motion
(after transforming to momentum space):
\begin{equation}
	\ddot{\chi} + 3H\dot{\chi}
+ \left[\frac{k^{2}}{R^{2}} + g^{2}b^{2} + 
g^{2}(a^{2}-b^{2}) \sin^{2}(m_{\phi}t)
\right]\chi = 0
\end{equation}
where $R$ is the scale factor of the universe ($H = \dot{R}/R$).
We can simplify this by using the substitution ${\rm X} \equiv R^{3/2}
\chi$. 
This gives:
\begin{equation}
	\ddot{{\rm X}} + \left[\frac{k^{2}}{R^{2}} + g^{2}b^{2}
+ g^{2}(a^{2}-b^{2})\sin^{2}(m_{\phi}t) -\frac{3}{4}H^{2}
-\frac{3}{2}\frac{\ddot{R}}{R} \right] {\rm X} = 0.
\end{equation}

Noting that the typical scale for $H$ is $m_{\phi} \ll a,b$ at the
time $\phi$ is coherently oscillating, we assume we can rewrite
this:
\begin{equation}
	\ddot{{\rm X}} + \left[\frac{k^{2}}{R^{2}} + g^{2}b^{2}
+ g^{2}(a^{2}-b^{2})\sin^{2}(m_{\phi}t)\right]{\rm X} =0.
\end{equation}
This is almost exactly the case of stochastic resonance in 
the expanding universe treated in \cite{linde2}.  The only
issue to be careful of is the size of $b$.  We will discuss
that shortly.
In the meantime, we will adapt section IX of 
\cite{linde2} to our purposes.

The fundamental procedure of the parametric resonance approach
is the rewriting of our equation of motion in the form of the
Mathieu equation:
\begin{equation}
	x''+\left(A-2q\cos(2z)\right)x = 0
\end{equation}
where prime denotes differentiation with respect to $z$
(please note that $q$ in this context is an unfortunate choice
for us -- it has nothing to do with quarks).
The appropriate substitutions here are:
\begin{equation}
	A = \frac{k^{2}}{R^{2}m_{\phi}^{2}} +
	\frac{g^{2}b^{2}}{m_{\phi}^{2}}
	+2q;
\end{equation}
\begin{equation}
	q = \frac{g^{2} (a^{2}-b^{2})}{4m_{\phi}^{2}};
\end{equation}
and
\begin{equation}
	z = mt.
\end{equation}

The work of \cite{linde2} shows that it is $q$ which determines
the efficiency of the condensate decay.  For $q > 10^{4}$, we expect
the condensate to retain a considerable $q$ value even after completion
of the ``first stage'' of preheating (with low
density of decay products).  If $q$ remains greater than order one
after this phase is completed we expect the decay to continue and
the condensate will give up a significant fraction of its energy
to decay products.  

We are interested, therefore, in estimating $q$ for a typical Q-ball
scenario.  At the beginning of
oscillations we expect $a^{2}-b^{2}$ to be of order $10^{20} m_{\phi}$
which gives a $q$ factor of:
\begin{equation}
	q \sim 10^{20} g^{2}.
\end{equation}
It is pointed out in \cite{linde2} that the $q = 10^{4}$ cutoff
value for strong preheating is weakly model dependent.  Given
that we expect $g^{2} \sim 0.1$ for strong interactions, however,
it seems safe to assume that we are 
well into the strong preheating regime (this is also true for the
$n=4$ flat directions, where we expect $q \sim 10^{14} g^{2}$).

Now let us consider, as we did for fermion production,
 what it would mean for this decay to 
go forward.  Once again, we see that the decay efficiency is proportional
to the difference $a^{2}-b^{2}$ so that it vanishes as we approach
a completely asymmetric condensate.  In fact, in the case of 
resonant production of bosons, we can have a second suppression
since $b$ is functioning as an effective bare mass in the formulas
above.  Thus, as mentioned in \cite{linde2} we might expect 
preheating to become inefficient if $2 b^{2} > a^{2}-b^{2}$.
It appears, then, that our decays take away the neutral condensate
but could leave a charged remnant.

The physics behind this result is straightforward.  The
four point coupling $g^{2}|\phi|^{2}|\chi|^{2}$
that we have considered here should 
only mediate annihilation, not true decays.  This would certainly
respect any baryon asymmetry present in the condensate.

This leaves us with an important possibility.  If decay into bosons
is strong enough, it could be that generic AD condensates will
damp much of their ellipticity.  This process could actually
aid Q-ball formation.  In fact, even the time scales estimated
in \cite{linde2} for the decay process are right for Q-ball formation.
They expect the first stage of resonant production to end at a
time scale of order  $100 m_{\phi}^{-1}$, just before Q-balls would form
according to numerical simulations \cite{KK}.  

There are two important issues to address, however, before we can say with
confidence
that annihilations help produce Q-balls.  First, it is important to note 
that we have not included
a study of rescattering of the produced $\chi$ particles.  Such 
back-reaction could have a negative impact on Q-ball formation,
and should be examined.
Second, annihilation would almost certainly have a negative
impact on 
Q-ball production in a very weakly charged condensate, 
since it could lead to
significant decay of the condensate 
before Q-ball formation would occur.  

\section{Conclusion}

In summary, decay of a partly charge-asymmetric Affleck-Dine condensate
into fermions will be strongly suppressed.  This suppression will give
more than enough time for the condensate to fracture into Q-balls.  
Annihilation of the neutral part of the condensate, however, can be
enhanced by non-perturbative effects completely analogous to preheating
in inflation.  These effects must be studied further, as they could
have important consequences for Q-ball formation.  In particular, it
seems that they should make Q-ball formation in strongly charged condensates
more likely, while suppressing formation in weakly charged condensates.

\section*{Acknowledgments}

The author wishes to thank A. Dolgov for constructive criticism, and 
G. Kane and F. Adams for their patience.

\end{document}